# Internal Dosimetry Assessment for Drinking the Groundwater of the Disi Aquifer

O. Nusair[a][1], W. Al-Tamimi[b], and O. Al-Qudah[c]

[a]Department of Physics and Astronomy, University of Alabama, Tuscaloosa, AL 35487, USA

[b]Department of Physics, Faculty of Science, The University of Jordan, Amman 11942, Jordan

[c]Jordan Atomic Energy Commission, P.O Box 70, Amman 11934, Jordan

## Abstract

The quest for a better understanding of the cancer risk associated with drinking the radium-contaminated groundwater of the Disi Aquifer in Jordan has become more urgent in recent years. To quantitively identify the health consequences attainable from the consumption of this groundwater source, internal dosimetry analysis was performed with emphasis on doses deliverable to bone surfaces. Moreover, the age-dependent dose calculations performed in this study show that the most critical group is those who are below the age of 15, where we predict an increase in the risk of cancer by up to a factor of 5 as compared to adults. It is also demonstrated that radium radioactivity remains relatively constant in the bone even 10 years after ingestion. The whole-body dose analysis concluded that it is a factor of 5 higher than what the WHO recommends as a limit.

## Keywords

Internal Dosimetry. Bone Surface Committed Dose Equivalent CDE. Disi Aquifer. Drinking Groundwater. Age-dependent Dosimetry. Radiation Protection.

---

[1] oynusair@ua.edu

## 1. Introduction

In the absence of a reliable freshwater resource in a heavily deserted country like Jordan, water scarcity can pose a serious threat to human lives. Being among the poorest countries in water supplies [1] a relatively poor quality of any available Jordanian groundwater source can be acceptable and justified by its benefits to the public. However, all aspects of the associated health risks should be carefully assessed and eventually addressed. Among those risks is what many studies [2-4] have found as a presence of elevated concentrations of the radionuclides $^{226}$Ra and $^{228}$Ra in the groundwater of the only reliable groundwater source, i.e. the Disi aquifer, in Jordan. In these studies, only whole-body doses attainable from drinking this radium-contaminated water were reported, and therefore the complete internal dosimetry analysis is needed. We attempt here to estimate the annual bone surface committed dose equivalent, in addition to other organs or tissues committed doses, based on the previously measured activity concentrations of $^{226}$Ra and $^{228}$Ra isotopes from [2] and [3]. Additionally, we provide an age-dependent dosimetric assessment.

Water collected from the Disi's fifty-five wells is brought to a collection reservoir located in the same well field. Therefore, radium concentrations measured for samples collected from this reservoir are the best to represent the overall quality of this groundwater source.

In Section 2, the general method of calculating internal doses based on a given intake is described. The annual effective doses are presented in Section 3 after a detailed analysis of organ related committed doses associated with drinking the Disi groundwater. Section 4 describes the differences in dosimetry between the six distinguishable age groups, highlighting the whole-body doses.

## 2. Calculations of $H_{T,50}$ and $H_{E,50}$

Internal dosimetry calculations are commonly performed to determine the dose deliverable to an organ or the whole body as a result of internal exposure to radiation. This exposure would deposit radionuclides in an organ or tissue, where the deposition of a given radionuclide is referred to as the fraction of uptake which enters the organ or tissue. The simple straight-through four-compartment model, introduced by the International Commission on Radiological Protection ICRP, is widely used to calculate these committed dose equivalents. As far as the ingestion of radionuclides is concerned, four defined sections of the gastrointestinal (GI) tract are modeled, namely: the stomach, the small intestine, the upper large intestine, and the lower large intestine. Using ICRP-30 methodology for calculating a committed dose equivalent, we attempted here to estimate a dose equivalent in target organ per disintegration of a radionuclide deposited in a source organ. However, when converting radiation exposure resulted from ingesting radium isotopes to internally delivered dose the limiting dose conversion coefficient is the bone surface coefficient ($h_{T,50}$) and not the effective coefficient for the whole body ($h_{E,50}$). Therefore, calculating Committed Dose Equivalent (CDE), or $H_{T,50}$, is of prime interest for examining the quality of an ingested substance that contains radium. Although many studies have found detectable quantities of the radionuclides $^{226}$Ra and $^{228}$Ra in elevated levels in the groundwater of the Disi aquifer in Jordan, almost none of these studies reported the important bone surface CDE. In this work, we estimated the annual bone surface CDE, in addition to other organs or tissues, based on previously measured activity concentrations of $^{226}$Ra and $^{228}$Ra isotopes. Two concentration measurements are used; one that was from nine water samples collected from the collection reservoir of the Disi aquifer, and the other was from 25 wells of the same aquifer.

The following formula is used to compute the tissue/organ annual committed dose equivalent received by adults from drinking the Disi water:

$$H_{T,50}\left(\frac{mSv}{yr}\right) = Intake\left(\frac{Bq}{yr}\right) \times Dose\ Conv.Coeff.\left(\frac{mSv}{Bq}\right) \quad (1)$$

Where $H_{T,50}$ is the committed dose equivalent to tissue or organ and is defined as the total dose equivalent deposited in the organ over the 50 years following the intake instant of the radionuclide. Some radionuclides are present in the body for short periods (weeks or even less), therefore the committed dose equivalent is regarded as a single contribution to the annual dose equivalent. In the case of the radionuclides that remain within the body indefinitely, due to a long physical half-life or a slow biological elimination, the dose equivalent is considered to accumulate at a constant rate over the person's lifetime.

To estimate the CDE (or $H_{T,50}$) from the ingested $^{226}$Ra and $^{228}$Ra activity concentrations (*A*) we use the ICRP-30 tabulated $h_{T,50}$ dose conversion coefficients for each target organ or tissue (Table 1) as follows:

$$H_{T,50}\left(\frac{mSv}{yr}\right) = A\left(\frac{Bq}{L}\right) \times h_{T,50}\left(\frac{Sv}{Bq}\right) \times 504\left(\frac{L}{yr}\right) \times 1000\left(\frac{mSv}{Sv}\right) \quad (2)$$

Where 504 L/yr is the annual drinking rate of an adult according to [5].

Similarly, the Committed Effective Dose Equivalent CEDE (whole-body dose) is given by:

$$H_{E,50}\left(\frac{mSv}{yr}\right) = A\left(\frac{Bq}{L}\right) \times h_{E,50}\left(\frac{Sv}{Bq}\right) \times 504\left(\frac{L}{yr}\right) \times 1000\left(\frac{mSv}{Sv}\right) \quad (3)$$

Where $h_{E,50}$ is the effective dose conversion coefficient. For the two radium isotopes of concern in this work, Table 1 illustrates these dose conversion coefficients [6].

Table 1 exposure-to-dose conversion coefficients for ingestion in ($\frac{Sv}{Bq}$). The limiting coefficients are indicated in bold

| Radionuclide | f₁ | Gonad | Breast | Lung | R Marrow | B Surface | Thyroid | Remainder | Effective |
|---|---|---|---|---|---|---|---|---|---|
| $^{226}$Ra | 0.2 | 9.16 10⁻⁸ | 9.17 10⁻⁸ | 9.16 10⁻⁸ | 5.98 10⁻⁷ | **6.83 10⁻⁶** | 9.15 10⁻⁸ | 1.03 10⁻⁷ | 3.58 10⁻⁷ |
| $^{228}$Ra | 0.2 | 1.58 10⁻⁷ | 1.57 10⁻⁷ | 1.57 10⁻⁷ | 6.53 10⁻⁷ | **5.82 10⁻⁶** | 1.57 10⁻⁷ | 1.63 10⁻⁷ | 3.88 10⁻⁷ |

## 3. H$_{T,50}$ and H$_{E,50}$ calculations associated with $^{226}$Ra and $^{228}$Ra radioactivity of the Disi groundwater

Bone seeking radionuclides, such as $^{226}$Ra and $^{228}$Ra, are commonly found in elevated concentrations in groundwater due to the presence of uranium and thorium in the rocks forming the walls of the host aquifer. The Disi aquifer in Jordan is no exception, where many studies reported the presence of detectable amounts of these isotopes in a variety of samples collected from extraction wells of the aquifer [2-4]. The average activity concentration of $^{226}$Ra and $^{228}$Ra from nine samples of groundwater from the main collection reservoir, reproduced from [2] in Table 2, is well representative of the Disi water and adequate for the internal dosimetry analysis in this work. $^{226}$Ra activity was measured in [2] using gamma-ray spectrometry with the direct detection of the $E_\gamma$= 186.2 keV peak, and $^{228}$Ra activity was measured by the detection of its daughter's $E_\gamma$= 911.2 keV line.

Radium ions can exchange with the calcium in a mineral medium, which makes bone a critical organ for radium [7-10]. In particular, the osteogenic cells in the bone surface are at risk of radiogenic cancer caused by radium ingestion. The ICRP-30 model contains detailed calculations of the dose equivalent to osteogenic cells, especially those on the endosteal surfaces of bone. These osteogenic cells are involved in the formation of osteoblasts (or new bone) and the resorption of osteoclasts (or old bone) and therefore are of concern to carcinogenesis in the bone.

Here we highlight the radium associated bone surface committed dose equivalent, using its conversion coefficients ($h_{T,50}$). The coefficients are given per unit of intake by ingestion of radionuclides $^{226}$Ra and $^{228}$Ra, expressed in (Sv Bq$^{-1}$) in Table 1 [6], as 6.83 x 10$^{-6}$ and 5.82 x 10$^{-6}$, respectively, with a 20% of fractional uptake from the small intestine to blood ($f_1$) for the common chemical forms of radium. However, this fractional uptake into the body fluids is the only parameter affected by the chemical form of the radionuclide. Using the committed effective dose equivalent conversion coefficients ($h_{E,50}$), which are 3.58 x 10$^{-7}$ and 3.88 x 10$^{-7}$ for $^{226}$Ra and $^{228}$Ra, respectively, one can estimate the combined Committed Effective Dose Equivalent (CEDE).

Table 2 indicates the annual CDE and CEDE values associated with a 1.4 L of daily drinking of water obtained directly from the collection reservoir by a member of the public.

**Table 2** Average $^{226}$Ra and $^{228}$Ra radioactivity concentrations in the collection reservoir (inherited from [2]) along with the estimated combined bone surface CDE (in bold) and other organ/tissue CEDE

| $^{226}$Ra concentration Bq/L | $^{228}$Ra concentration Bq/L | B Surface CDE mSv/yr | Gonad mSv/yr | Breast mSv/yr | Lung mSv/yr | R Marrow mSv/yr | Thyroid mSv/yr | Remainder mSv/yr | CEDE mSv/yr |
|---|---|---|---|---|---|---|---|---|---|
| 0.44 ± 0.08 | 1.05 ± 0.03 | **4.60 ± 0.29** | 0.10 ± 0.01 | 0.10 ± 0.01 | 0.10 ± 0.01 | 0.48 ± 0.03 | 0.10 ± 0.01 | 0.11 ± 0.01 | 0.29 ± 0.02 |

The total radioactivity of radium isotopes that remain in the body after ingestion of radium with the specified dose of 0.29 mSv in the first year is plotted below as a function of time for the next 50 years, as shown in Figure 1. It is clear from the figure that the skeleton radioactivity content (sum of the contents in the bone surface, bone volume, red marrow and skeleton blood fraction) comprises the majority of the radioactivity content in the whole body and this content remains relatively constant for the following 10 years after ingestion. Whereas the liver content of radioactive radium clears after 10 days after ingestion.

**Figure 1 Time dependence of the $^{226}$Ra and $^{228}$Ra radioactivity content (Bq) in an organ associated with a Committed Effective Dose Equivalent (CEDE) of 0.29 mSv.**

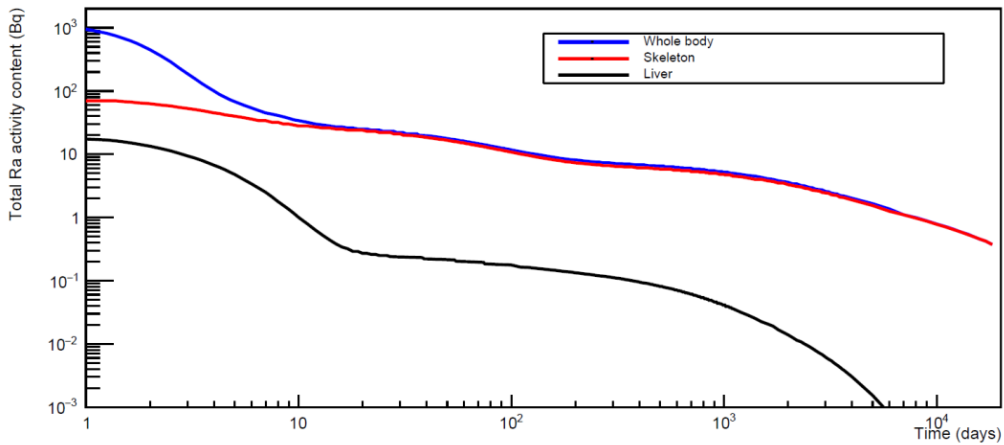

From another study published by [3], we averaged the activity concentrations of $^{226}$Ra and $^{228}$Ra from the thirteen wells in the unconfined part of the aquifer with the concentrations from the twelve wells in the confined part of the aquifer to get 0.60 Bq L$^{-1}$ and 1.46 Bq L$^{-1}$, for $^{226}$Ra and $^{228}$Ra, respectively. The internal dosimetry analysis results, found in Table 3, are similar to those obtained based on the concentrations measured by [2].

**Table 3** Average $^{226}$Ra and $^{228}$Ra radioactivity concentrations from [3] along with the estimated organ/tissue CDEs and combined CEDE

| $^{226}$Ra concentration Bq/L | $^{228}$Ra concentration Bq/L | B Surface CDE mSv/yr | Gonad mSv/yr | Breast mSv/yr | Lung mSv/yr | R Marrow mSv/yr | Thyroid mSv/yr | Remainder mSv/yr | CEDE mSv/yr |
|---|---|---|---|---|---|---|---|---|---|
| 0.60 | 1.46 | **6.35** | 0.14 | 0.14 | 0.14 | 0.66 | 0.14 | 0.15 | 0.39 |

## 4. Age-dependent $H_{E,50}$

In the ICRP's Publication 23 a Reference Man is defined as a young Caucasian male, 20-30 years old, weighing 70 kg and is 170 cm tall. This category fits most people who drink from the Disi Aquifer in Jordan. However, aging was found to significantly decrease radium absorption by a factor of 2-4 when compared to adults [11], and therefore it is important to take it into account when estimating the internal dose associated with radioactive radium ingestion. The ICRP identifies six different age categories ranging from 3 months to adulthood as indicated in Table 4 below.

Age-dependent dose coefficients are found in Publication 72 [12], which were calculated using the ICRP-30 model of the GI tract. In this model dose coefficients for radioisotopes don't take into consideration the biokinetics of organ retention and excretion following absorption into the blood. However, as stated in Publication 72, the calculation of children's dose coefficients using the adult biokinetic parameters will tend to overestimate doses. One reason for this is that rates of elimination from tissues and excretion are generally greater at younger ages. To overcome this problem, the biokinetic models were developed to consider age dependence.

Table 4 illustrates the results of this age-dependent dose analysis based on radium concentrations from [2]. It is noted that the young groups are more affected by drinking the groundwater of the Disi aquifer than adults. The whole-body dose is about 5-23 times higher than the recommended WHO dose limit [13] and 1-5 times higher than the Jordanian limit [14].

**Table 4 Age-dependent whole-body dose based on coefficients from [15] and annual water drinking rates from [5]**

| Age Category | Dose Conversion Coefficients for ingesting $^{226}$Ra (Sv/Bq) [15] | Dose Conversion Coefficients for ingesting $^{228}$Ra (Sv/Bq) [15] | Water Drinking rate (L/yr) [5] | CEDE (mSv/yr) |
|---|---|---|---|---|
| Infant | 4.7E-6 | 3.0E-6 | 69 | 2.32 ± 0.07 |
| 1 year | 9.6E-7 | 5.7E-6 | 80 | 0.50 ± 0.02 |
| 5 years | 6.2E-7 | 3.4E-6 | 190 | 0.73 ± 0.02 |
| 10 years | 8.0E-7 | 3.9E-6 | 250 | 1.11 ± 0.03 |
| 15 years | 1.5E-6 | 5.3E-6 | 294 | 1.83 ± 0.06 |
| Adult | 2.8E-7 | 6.9E-7 | 504 | 0.43 ± 0.02 |

## 5. Conclusions

In this study, we reported organ/tissue committed doses and whole-body dose associated with drinking the groundwater of the Disi Aquifer in Jordan. The study extended previous efforts that estimated radiological doses and added to them organ doses and age-dependent whole-body doses. Based on the findings from this work we recommend that the national agencies in Jordan should maintain close surveillance over all wells which feed the collection reservoir to control the collective total effective dose. Rationally speaking, it can be justified by its benefits to the public that dose contributions from the drinking of the Disi groundwater to the total population dose as estimated in this study are acceptable.

## 6. References

1. Haddadin N, Qaqish M, Akawwi E, Bdour A (2010) Water shortage in Jordan-sustainable solutions. Desalination 250: 197-202
2. Al-Qararah, A., Al-Qudah, O., Alameer, S. *et al.* High radioactivity levels of radium isotopes in groundwater of the Disi aquifer. *J Radioanal Nucl Chem* **322,** 1995–2001 (2019).

3. Vengosh A, Hirschfeld D, Vinson D, Dwyer G, Raanan H, Rimawi O, Al-Zoubi A, Akkawi E, Marie A, Haquin G, Zaarur S, Ganor J (2009) High Naturally Occurring Radioactivity in Fossil Groundwater from the Middle East. Environ. Sci. Technol.43: (1769–1775).

4. Dababneh S (2014) Comment on "High Naturally Occurring Radioactivity in Fossil Groundwater from the Middle East". Environ. Sci. Technol.48, (9943−9945).

5. Ershow A.G. Cantor K.P. (1989). Total water and tap water intake in the United States: Population-based estimates of quantities and sources. Bethesda, MD: FASEB/LSRO.

6. Environmental Protection Agency (EPA). (1988). *Limiting values of radionuclide intake and air concentration and dose conversion factors for inhalation, submersion, and ingestion* (Federal Guidance Report No. 11).

7. Finkelstein M M (1994) Radium in drinking water and the risk of death from bone cancer among Ontario youths. Can. Med. Assoc. J. 5:(565–571).

8. Finkelstein M M, Kreiger N (1996) Radium in drinking water and risk of bone cancer in Ontario youths: A second study and combined analysis. Occup. Environ. Med. 53:(305–311).

9. Mays C W, Rowland R E, Stehney A F (1985) Cancer risk from the lifetime intake of Ra and U isotopes. Health Phys. 48: (635–647).

10. Cohn P, Skinner R, Burger S, Fagliano J, Klotz J (2003) Radium in Drinking Water and the Incidence of Osteosarcoma. New Jersey Department of Health and Senior Services: Trenton, NJ. 17 pp.

11. Taylor D M, Bligh P H, Duggan M H (1962). The absorption of calcium, strontium, barium and radium from the gastrointestinal tract of the rat. *Biochem J*. 83(1):25-29.

12. Age-dependent doses to members of the public from intake of radio-nuclides (Part 5). Compilation of ingestion and inhalation coefficients. ICRP Publication 72. Ann. ICRP 26(1).

13. WHO (2011) Guidelines for drinking-water quality, 4th edn. WHO, Geneva.

14. Technical Regulation (Mandatory) 286/2008 on water-drinking water (2008). Jordan standards and metrology organization, Amman.

15. ICRP (2012) Compendium of Dose Coefficients based on ICRP Publication 60. ICRP Publication 119. Ann. ICRP 41(Suppl.).